\documentclass{svjour2}                    
\smartqed  
\usepackage{graphicx}
\journalname{J Stat Phys}
\begin{document}

\title{Close relationships: A study of mobile communication records
}

\author{Vasyl Palchykov       \and                       
             J\'anos Kert\'esz      \and 
             Robin Dunbar		\and 
             Kimmo Kaski                   
}

\institute{V. Palchykov \at
              Department of Biomedical Engineering and Computational Science (BECS), Aalto University School of Science, P.O. Box 12200, FI-00076, Finland.
             \emph{Permanent address:} Institute for Condensed Matter Physics, National Academy of Sciences of Ukraine, UA-79011 Lviv, Ukraine\\  
           \and
           J. Kert\'esz \at
             Center for Network Science, Central European University, Nador u. 9, H-1051 Budapest,
             Hungary Institute of Physics BME, Budapest, Budafoki ut 8., H-1111, and
             BECS, Aalto  University School of Science, P.O. Box 12200, FI-00076, Finland\\
           \and
          R.I.M. Dunbar \at
          Department of Experimental Psychology, University of Oxford, South Parks Road, Oxford OX1 3UD, UK.
           \and
           K. Kaski \at
           BECS, Aalto  University School of Science, P.O. Box 12200, FI-00076, Finland\\
              \email{kimmo.kaski@aalto.fi}           
}

\date{Received: date / Accepted: date}

\maketitle

\begin{abstract}
Mobile phone communication as digital service generates ever-increasing datasets of human communication actions, which in turn allow us to investigate the structure and evolution of  social interactions and their networks. These datasets can be used to study the structuring of such egocentric networks with respect to the strength of the relationships by assuming direct dependence of the communication intensity on the strength of the social tie. Recently we have discovered that there are significant differences between the first and further "best friends" from the point of view of age and gender preferences. Here we introduce a control parameter $p_{\rm max}$ based on the statistics of communication with the first and second "best friend" and use it to filter the data. We find that when $p_{\rm max}$ is decreased the identification of the "best friend" becomes less ambiguous and the earlier observed effects get stronger, thus corroborating them.
\end{abstract}

\section{Introduction}\label{intro}
Information Communication Technology (ICT) has been (and still is) 
providing a ple\-thora of new services for every individual in society to use, which then means that every user transaction is recorded as a kind of "digital footprint". These footprints of ours form ever-growing datasets, such as those of mobile phone communication (MPC) of operators, the access to which can open up  quite unparalleled views on social interactions between individuals on a societal scale and, in general, to the structure and dynamics of the society \cite{reviews}. So studying these datasets through computational analysis and modeling gives us insight to the system of social interactions \cite{Eagle09}, which is different from, but at the same time complementary to, that based on questionnaires. The main differences are that the latter approach is wide in its social scope yet having quite ill-defined and subjective scale of social closeness, while the former, e.g. that of MPC, is narrow in its social scope consisting only  two communication channels, voice calls and text messages (SMS), measured by their frequencies or in case of calls also by duration.  There are by now some good examples of what can be learned and understood about these systems and their behavior by using modern data and reality mining methods \cite{Eagle06} as well as other computational analysis and modeling approaches \cite{Onnela07}. 

The closeness of social relationship is one of the fundamental issues in sociology and it is crucial in the extended social brain hypothesis \cite{Dunbar98}, which states that the capacity of forming social ties is limited to 150 by the brain capacity due to human evolution. Within this "Dunbar circle" further ordering takes place. The number of very close relationships is very small (3-5) \cite{Hill03}, and in order to be able to maintain them a new close relationship unavoidably tends to make an older one weaker \cite{Saramaki12}.

In a recent study \cite{Palchykov12} we have analyzed a seven months period of mobile phone records of 3.2 million subscribers from a European provider. We show that there are striking differences between the gender preferences of the "best friend" (defined by the most frequently contacted partner in the egocentric network) and relationships of lesser ranks. The best friend is found to be of the opposite sex for ages less than 50 years with a peak around 30. The peak for females is higher and appears earlier or at a little younger age than for males. Similarly a higher female focus on close opposite-sex relationships than the focus of men has been observed for the text message communication in online social networks \cite{Backstrom11}.
Above 50 there is a further difference between males and females: While males still prefer females as best friends, females have a slight preference to the same gender. The situation is very different for the second and further best friends. Below 50 gender homophily can be observed, but above that age males tend to be gender neutral in their selection, while females have a tendency to select males. These observations have anthropological interpretation in terms of gender dependent reproductive investment and they demonstrate that our actions are largely motivated by basic evolutionary aspects even for the usage of ICT.

The weak point in the above reasoning is that the ranking of "friends" is taken as uniquely determined by the order of MPC intensities or the frequency of contacts. While communication intensity in general could be a good measure of social tie strength \cite{Granovetter,Roberts09,Roberts11} MPC offers only two channels of communication among many other possibilities, like face to face, email, social network sites etc. Also there are person to person variations with respect to the usage of these channels. By the same token the set of close relationships for an ego identified on the basis of frequency of calls is not always the set of friends an ego spends the most time with in communication. Indeed, studying MPC records (throughout this paper we use the same dataset as in  \cite{Palchykov12}) we show that different ways to determine the strength of a social tie may suggest different persons as the best friends. This raises the question of reliability for to identify the closest relationships. In this study we introduce a simple parameter and show that it allows us to single out the egos with reliably identified best friend and thus to control the corresponding reliability level. Our analysis shows that the more reliably identified the best friend the more clearly his or her privileged status is observed and the more pronounced gender differences in communication are found.

Note that the dataset under investigation contains all the communication records of each ego subscribing the given service provider independent of whether the ego's friends are subscribers of the same provider or of a different one. If the ego's friends subscribe a service provider different from that of an ego, then the age and gender information of the friends is not available but the remaining information allows us to rank all friends, which is particularly important for the analysis in this study.

After this introduction we present a section on the identification of the close relationships and sections regarding gender preferences and age correlations. Then we end with the section of conclusions. 

\section{Filtering of the closest relationships}
The intensity of communication may have different measures depending on the communication channel. Even within one channel there could be several characteristics such as for phone calls the number of (outgoing or bidirectional) calls and (outgoing or bidirectional) call durations. This makes the determination of the strength of social ties or social closeness difficult even within the assumption that there is a direct relationship with the intensity of communication. Indeed, beside an overall positive correlation between the number of calls and call durations \cite{Onnela07a}, our results show that the strongest relationship, identified on the basis of the number of calls and on the basis of duration of calls coincides for about two of three egos, but for every third ego these definitions give different results.

Here we first assume that the number of calls by an ego to his or her friends serves as a measure of the strength of these relationships. Then the best friend of an ego is defined as the alter whom the ego calls most frequently, the second best friend as the alter whom the ego calls the second most frequently and so on for the third, fourth etc. best friend. This means that the order of the closeness of relationships for each ego is defined by the number of outgoing calls. 

As the next step we consider how reliably is the best friend identified, which is an open question. Indeed, let us focus on an ego $i$ and assume that he or she made $n_1$ calls to the best friend and $n_2$ calls to the second best friend. If $n_1$ is large enough and is considerably larger than $n_2$, then it is natural to suggest that the best friend is reliably identified. However, if $n_1$ and $n_2$ have similar values or these values are small enough, the reliability to identify the best friend will decrease.

In order to estimate the reliability level for ego $i$ we focus on his or her two strongest ties and consider a no-preference null model as described below. Assume that there are no preferences for the ego $i$ between his or her best friend and second best friend. Then the difference between the numbers $n_1$ and $n_2$ is of purely fluctuational or random origin. This case may be modelled by assuming that ego $i$ distribute $n = n_1+n_2$ calls randomly over two closest friends. Then for each ego $i$ we assign p-value $p_i$ defined as the probability that as a result of described null model the ego makes at least $n_1$ calls to his or her the best friend, and not more than $n_2$ calls to the second best friend.
The probabilities of each particular realization are binomially distributed and then the p-value $p_i$ is
  \begin{equation}
    p_i = \frac{1}{2^n}\sum_{x_1=n_1}^n \frac{n!}{x_1! (n-x_1)!}.
  \end{equation}
Here $x_1$ runs over all the allowed number of calls to the best friend and the factor $1/2$ gives the probability for each particular call to be directed to one of ego's friends.
The p-value $p_i$ varies within a range $p_i\in[0,1]$ , $i = 1,\ldots,N$. If the p-value is high enough ($p_i\to 1$), then the observed configuration ($n_1$, $n_2$) occurs with a high probability thus being explainable by no-preference model. However, for small $p_i$ ($p_i\to0$) it is unlikely that the observed configuration (or even more extreme configurations) 
may be explained by a simple no-preference model, hence, the choice of the best friend indicates ego's real preference.

\begin{figure}[!h]
\includegraphics[width=0.45\textwidth]{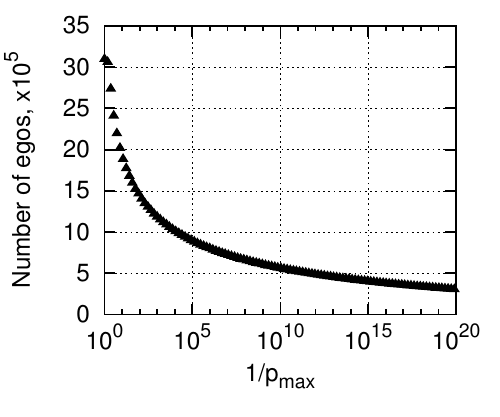}
\caption{Number of subscribers (egos) whose communication patterns satisfy the condition $p_i\leq p_{\rm max}$ as a function of $p_{\rm max}^{-1}$. Note that although this filtering procedure reduces statistics, there still remains more than 500 000 egos left even for $p_{\rm max} = 10^{-10}$.}
\label{fig1}
\end{figure}

\begin{figure}[!h]
\includegraphics[width=0.9\textwidth]{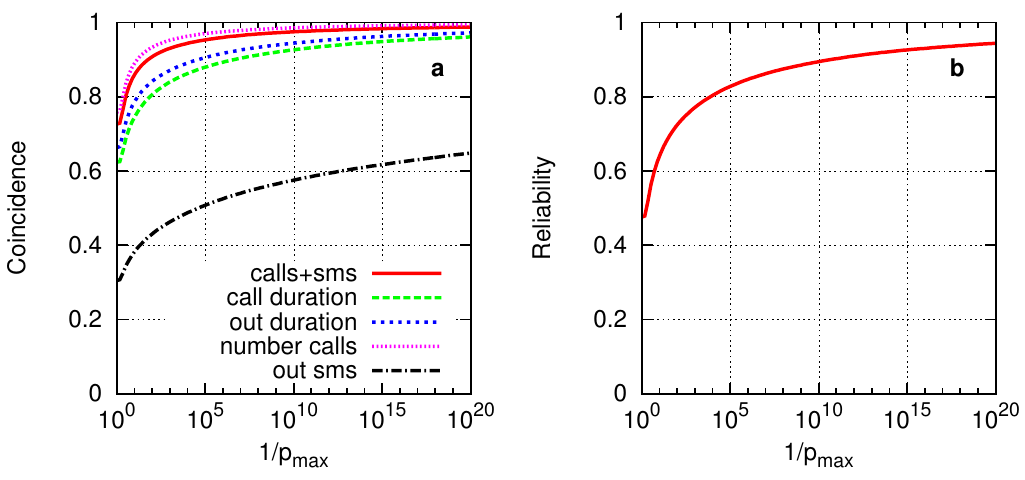}
\caption{Coincidence between different definitions to identify the best friend of an ego and its reliability. Panel {\bf a}: probability that the best friend defined on a basis of the number of outgoing calls with given value of $p_{\rm max}$ coincides with the best friend defined by using other five different criteria: number of calls and text messages over a link (red line), total duration of calls with the friend (green line), duration of outgoing calls (blue line), total number of calls over a link (purple line) and number of text messages sent by an ego (among those egos who used text messages at all) (black line) as a function of $p_{\rm max}^{-1}$. Apart from using only text messages in the definition all other curves coincide quite closely. Panel {\bf b}: reliability of identifying the best friend for given value of $p_{\rm max}$, defined as the probability that the five different ways to single out the best friend, i.e. based on the number of out-calls, number of calls over a link (in + out), duration of out-calls, total duration of calls, and number of calls and text messages over a link, lead to the same alter as the best friend of an ego as a function of $p_{\rm max}^{-1}$.}
\label{fig2}
\end{figure}

Below we show that the p-value allows us to filter out the egos with reliably identified the best friend such that it can be 
used as a control parameter for our analysis. 
For this reason we establish the threshold value $p_{\rm max}$ and considering the egos with $p_i\leq p_{\rm max}$ in order to achieve a higher reliability level for the best friend identification. Of course, this filtering reduces the sample size (as depicted in Fig.~\ref{fig1}), but not too much even for the small values of $p_{\rm max}$, for sufficient statistics. 
In Figs.~\ref{fig2}{\bf a} and {\bf b} we show the results of comparisons between five different definitions for to identify the best friends of the egos (see the figure caption for details). Here we see a clear increase in the percentages of egos for whom the different definitions lead to the same individual as the best friend. In particular, in Fig.~\ref{fig2}{\bf a} we see that by setting the value of the control parameter $p_{\rm max} = 10^{-5}$, the fraction of remaining egos for whom both the frequency and duration of calls lead to the same alter as the best friend, increases up to about $90\%$, thus reducing the fraction of ill-defined cases by a factor of $3$. 
Decreasing the value of the control parameter $p_{\rm max}$ below $10^{-5}$, the fraction of ill-defined cases  reduces even further, thus leading to very high degree of coincidence between all but one of the definitions the best friend alter for an ego, namely the definition using only text messages. It is indeed evident in Fig.~\ref{fig2}{\bf a} that there is comparatively low level of overlap or coincidence between voice calls and text messages as separate channels of communication. This may be caused by these two channels serving different functions in human communication. In Fig.~\ref{fig2}{\bf b} we show the relative reliability of identifying the best friend alter of an ego as a function of the control parameter $p_{\rm max}$ for five different definitions all including voice calls. These five curves overlap indistinguishably, thus leading to same alter as the best friend with very high degree of reliability.   

We can conclude here that a single parameter $p_{\rm max}$ allows us to control the level of reliability in the identification of the best friend by selecting those subscribers for whom several ways to define the closest relationship lead to the same person. In the next section \ref{sect_gender} we consider whether the patterns observed in \cite{Palchykov12} are maintained with this change of reliability level. We also investigate the gender differences in mobile communication.

\section{Gender differences}\label{sect_gender}

It has been shown \cite{Palchykov12} that according to gender correlations only one of all the friends has a special status: only the closest relationship demonstrates a clear tendency to be characterized by the opposite gender to that of an ego. This bias turned out to be highest during the reproductively active period. All other friends who are of lower ranks or less close show considerably smaller gender bias towards the ego's own gender. 

\begin{figure}[!h]
\includegraphics[width=0.95\textwidth]{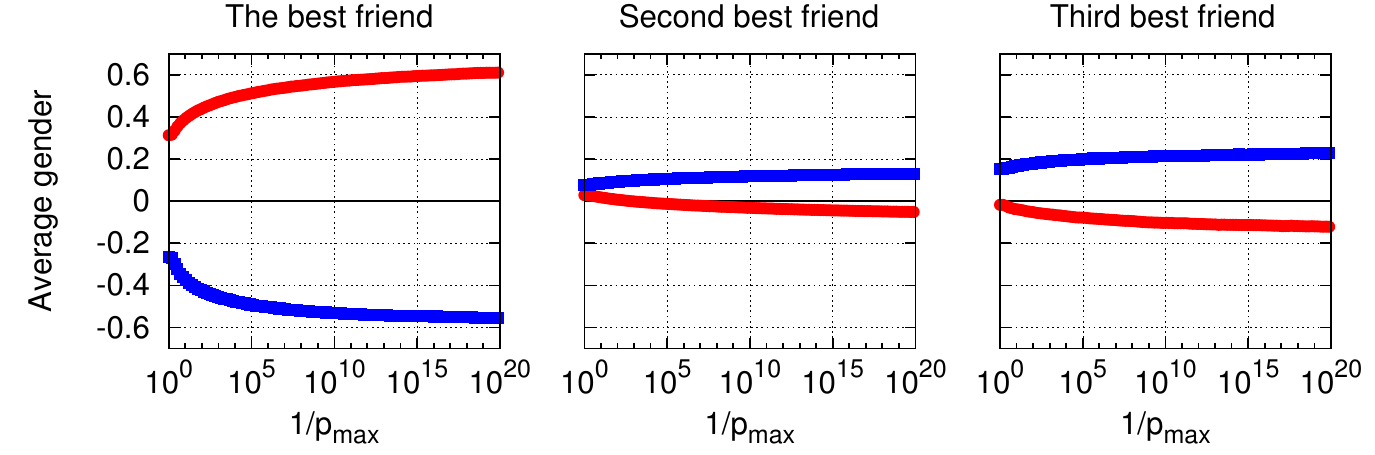}
\caption{Average gender of the best friend (panel {\bf a}), second best friend (panel {\bf b}) and the third best friend (panel {\bf c}) as a function of $p_{\rm max}^{-1}$ for males (blue squares) and females (red balls). The special status of the best friend is demonstrated by the strong bias towards the opposite gender, which is  in contrast to less bias in the same gender relationships for the second and third best friends. The strengthening of the opposite gender preference for best friend as a function of the inverse control parameter  ($p_{\rm max}^{-1}$) verifies our previous findings \cite{Palchykov12}.
}
\label{fig3}
\end{figure}

In order to analyze the evolution of this relationship with $p_{\rm max}$ we assign a gender variable $g_i$ for each subscriber $i$ in such a way that $g_i = \pm 1$ for male and female subscribers, respectively. Defining $f_i$ to be a gender of the best friend for subscriber $i$, we get the average gender $\langle{f}\rangle$ of the best friend as follows:
\begin{equation}
\langle{f}\rangle = \frac{\sum_i f_i}{\sum_i 1}.
\end{equation}
Here $i$ runs over all egos, subject to given restrictions. If the restriction is taken conditional to gender of the egos, we get 
the average gender of the best friend for males and females, separately. Then if we include  the threshold $p_{\rm max}$ as an additional restriction, our original finding \cite{Palchykov12} about the special status of the best friend of an ego gets corroborated. This is clearly seen in Fig.~\ref{fig3}, where we show the average gender of the best friend, second best friend, and the third best friend as a function of $p_{\rm max}^{-1}$. 

In order to look more deeply into the gender preferences and to check our previous findings about the gender dependent differences in the reproductive investment, we consider the average gender of close relationships as a function of the age and gender of the ego.

\begin{figure}[!h]
\includegraphics[width=0.95\textwidth]{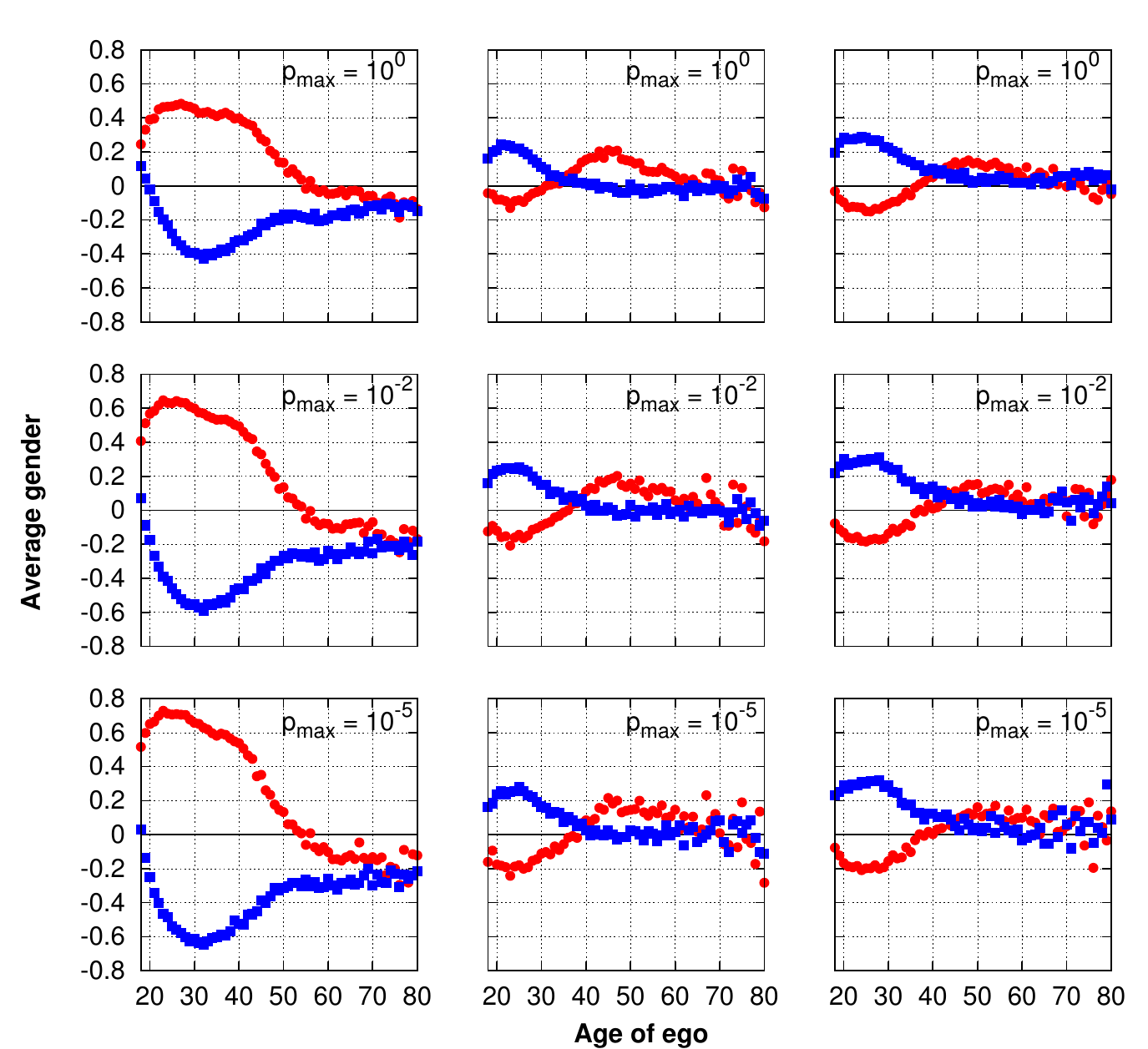}
\caption{Average gender of the best friend (left panel column), second best friend (central panel column) and the third best friend (right panel column) as a function of the age of an ego for different values of the control parameter $p_{\rm max}$. Blue squares correspond to male egos and red balls to female egos.}
\label{fig4}
\end{figure}

The average gender of the best friend, second best friend, and the third best friend as a function of the age of ego are shown in Fig.~\ref{fig4} for males and females separately, showing high level of gender-age correlations.
While there is no significant dependence in the curves describing the second and thirds best friends as a function of $p_{\rm max}$, the special status of the best friend becomes again apparent. In fact, we observe that i) females have a stronger bias towards males during their reproductive period than the males towards females; ii) this changes at around the age of 50 when there is some bias for males but not for females. The effects becomes stronger with decreasing $p_{\rm max}$, supporting our pervious findings obtained without the threshold.

\begin{figure}[!h]
\tabcolsep 2mm
\begin{tabular}{cc}
\includegraphics[width=0.45\textwidth]{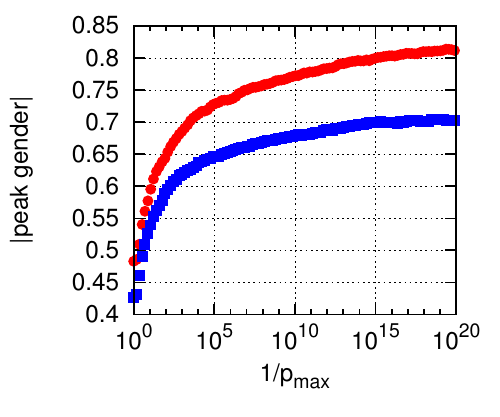}&
\includegraphics[width=0.45\textwidth]{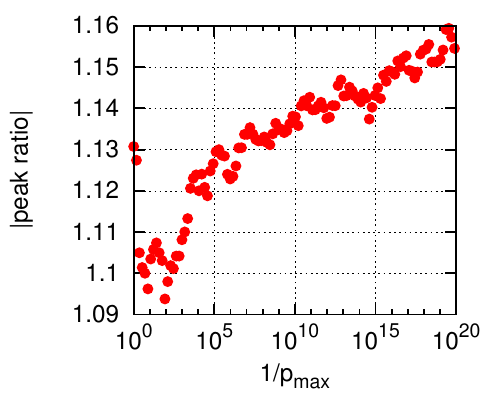}\\
\end{tabular}
\caption{Panel {\bf a} shows the highest (absolute) value of the average gender of the best friend, for males (blue squares) and females (red circles). Panel {\bf b} shows the ratio between these values for females and males as a function of the inverse control parameter $p_{\rm max}^{-1}$. The increasing tendency of this ratio corroborates our previous finding that females are more focussed on opposite sex relationships than males during their reproductively active period.}
\label{fig5}
\end{figure}

In more quantitative terms,  for $p_{\rm max} = 1$ the highest absolute value for the best friend gender for female $|\langle{f}\rangle|\approx0.48$ (meaning 74 males as the best friends among 100) exceeds the corresponding value for males ($|\langle{f}\rangle|\approx0.43$). The change with $p_{\rm max}$ of the highest absolute value of the best friend gender bias is shown in Fig.~\ref{fig5}{\bf a}. It shows that the effect of opposite sex relationships becomes more evident both for males and females if the best friend is defined more reliably. Moreover, Fig.~\ref{fig5}{\bf b} giving the ratio between peak values for females and males demonstrates that the difference between males and females becomes more pronounced with decreased $p_{\rm max}$.

As a result we can see that using a threshold $p_{\rm max}$ for controlling the reliability of identifying the best friend actually improves our earlier results about the gender preference in selecting the best friend as well as the special status of the ego having for opposite gender.  Similarly the very different strategies for reproductive investment for males and females became even stronger when we decrease the value of $p_{\rm max}$. 

Apart from the role gender plays in choosing the best friend alter of an ego, also age plays an important role for choosing the set of close relationships \cite{Palchykov12}.  In the next section \ref{sectionAge} we will consider the corresponding effect of age and will examine how the results obtained for gender preferences and their changes with $p_{\rm max}$ are reflected as changes in the age structure of close friendships.

\section{Age correlations}\label{sectionAge}
\begin{figure}[!h]
\tabcolsep 2mm
\includegraphics[width=0.50\textwidth]{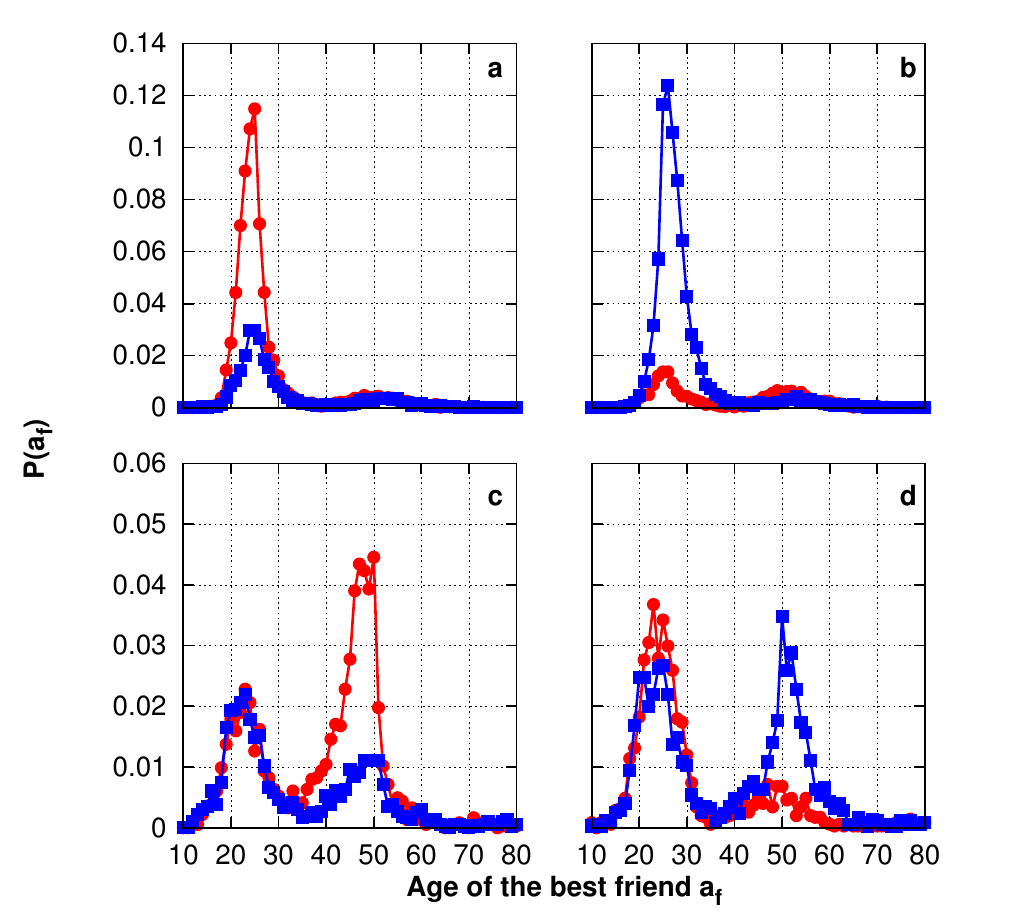}
\caption{Age distribution of the best friend for 25 years old male egos (panel {\bf a}), 25 years old female egos (panel {\bf b}), 50 years old male egos ({\bf c}), and 50 years old female egos ({\bf d}). Blue squares and red balls correspond to male and female best friends, respectively. Here the control parameter is set $p_{\rm max} = 10^{-2}$.}
\label{fig6}
\end{figure}

In our earlier study \cite{Palchykov12} we investigated the influence of the ego's age on the age distribution of the best friends. A strong effect was observed, such that while the distribution turned out to be always bimodal with a distance between the peaks corresponding to the generation gap, the peaks corresponding to the parents generation was much smaller in the case of the young people than the other way around. Moreover, females older that 50 years of age had a preference towards females of the young generation, possible their daughters. This led us to conclude that there is a shift in the reproductive investment strategy of females towards the daughters (and through them, towards the grandchildren) after the onset of the menopause. 
In Fig.~\ref{fig6} the distribution of best friend's age for $p_{\rm max} = 10^{-2}$ is shown  for the cases when the ego is 25 and 50 years of age.
This figure shows that independently of the age and gender of the ego the strong bias toward the opposite-sex relationships appears around ego's own age, but for the age difference of generation gap this bias for opposite sex is more balanced. Nonetheless, for 50 years of age females there is a significant preference towards females (daughters).  The bi-modality of this distribution and different gender preferences around each mode leads us to suggest different dependence of these relationships on $p_{\rm max}$.

\begin{figure}[!h]
\tabcolsep 2mm
\includegraphics[width=0.95\textwidth]{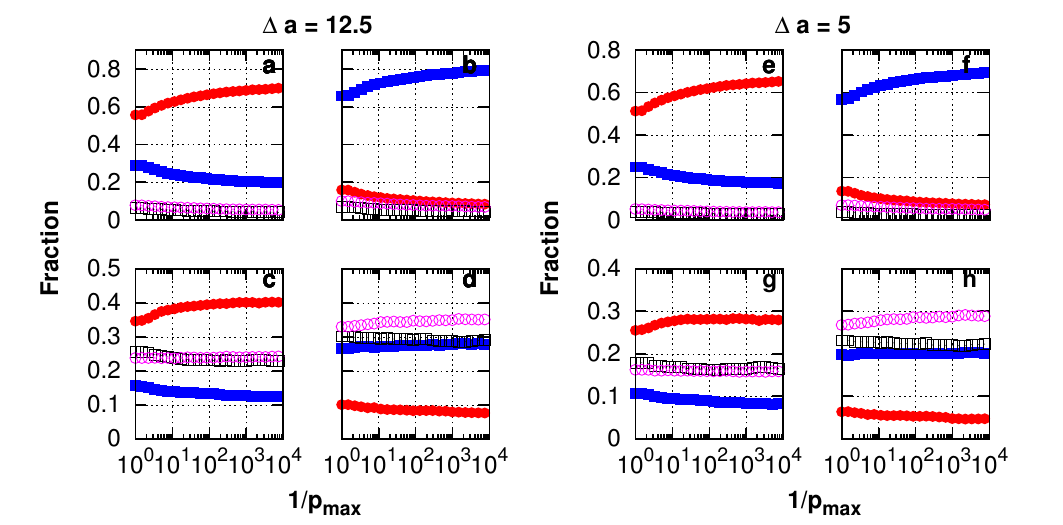}
\caption{Fraction of male best friends (blue squares) and female best friends (red balls) that differ in age from the ego not more than $\Delta a$ for different values of the control parameter $p_{\rm max}$. Purple circles  and black squares provide the fraction of female and male best friends, respectively, whose age differs from ego's own age by $25\pm\Delta a$. On the left hand side panels ({\bf a}, {\bf b}, {\bf c} and {\bf d}) $\Delta a = 12.5$ while on the right hand side panels ({\bf e}, {\bf f}, {\bf g} and {\bf h}) $\Delta a  = 5$. Panels {\bf a} and {\bf e} corresponds to 25 years old male egos, {\bf b} and {\bf f}: 25 years old female egos, {\bf c} and {\bf g}: 50 years old male egos, {\bf d} and {\bf h}: 50 years old female egos. }
\label{fig7}
\end{figure}

In order to investigate the evolution of the age structure of close friends we will take the following approach. First, we single out those best friends whose age $a_f$ is similar to the age $a$ of an ego $a_f\in[a-\Delta a, a+\Delta a]$, where $\Delta a$ defines the width of age interval. Then we select those best friends whose age correspond to a parent-child relationship with the same width of the age interval $a_f\in[a-25-\Delta a, a-25+\Delta a] \cup [a+25-\Delta a, a+25+\Delta a]$. In Fig.~\ref{fig7} we  show how the fraction of best friends whose age is either similar to ego's own age or differs by about a generation varies with $p_{\rm max}$ for $\Delta a = 12.5$ and $\Delta a = 5$.
This figure shows clearly that only the fraction of the best friends of similar age and opposite gender to that of the ego increases with $p_{\rm max}^{-1}$. This tendency is more clearly observed for younger subscribers, which seems to agree with our findings of gender preferences in section \ref{sect_gender}.

\section{Conclusions}
In this study the problem of identifying the closest relationships within a social network based on a large MPC dataset has been considered. This problem is challenging, both because in reality human communication is multi-channeled varying from person to person and because the MPC dataset gives us scope of only two channels of communication with measurable features like frequency and duration. In this study we have shown that the ranking order of close relationships defined by using the frequency of calls does not always coincide with the ranking order defined by using the duration of calls as the measure of closeness in the relationship. In order to overcome this problem we defined a single additional parameter (p-value), which allowed us to effectively single out those subscribers for whom the closest relationships can be reliably identified.
Increasing the level of reliability in identifying the best friend by restricting egos with the control parameter $p_{\rm max}$  
we corroborated the previously obtained results, namely, the special status of the "best friend" and the gender differences in communication with him or her. These findings are also reflected as changes in the age structure of close relationships. 

\begin{acknowledgements}
We thank A.-L.~Barab\'asi for providing access to the dataset used in this research. Financial support from EU's 7th Framework Program FET-Open to ICTeCollective project no. 238597 and by the Academy of Finland, the Finnish Center of Excellence program, project no. 129670, and TEKES (FiDiPro) are gratefully acknowledged. RIMD is supported by a European Research Council Advanced grant.
\end{acknowledgements}

\end{document}